\documentclass[13 pt]{article}
\usepackage{amssymb}

\def\om{\omega}

\def\non{\nonumber}
\def\t{\tilde}

\newcommand{\beq}{\begin{equation}}
\newcommand{\eeq}{\end{equation}}
\newcommand{\bea}{\begin{eqnarray*}}
\newcommand{\eea}{\end{eqnarray*}}
\newcommand{\beqa}{\begin{eqnarray}}
\newcommand{\eeqa}{\end{eqnarray}}

\begin{document}
\newfont{\elevenmib}{cmmib10 scaled\magstep1}%

\newcommand{\preprint}{
            \begin{flushleft}
            \end{flushleft}\vspace{-1.3cm}
            \begin{flushright}\normalsize  \sf preprint-number\\
            MISC--2010--08\\
            \end{flushright}}
\newcommand{\Title}[1]{{\baselineskip=26pt \begin{center}
            \Large   \bf #1 \\ \ \\ \end{center}}}
\hspace*{2.13cm}%
\hspace*{1cm}%
\newcommand{\Author}{\begin{center}\large
          Kozo Koizumi\footnote{kkoizumi@cc.kyoto-su.ac.jp}
\end{center}}
\newcommand{\Address}{{\baselineskip=18pt \begin{center}
          \it Maskawa Institute for Science and Culture\,(MISC),\\
           Kyoto Sangyo University, Kyoto 603-8555, Japan
      \end{center}}}
\baselineskip=13pt

\preprint
\bigskip

\Title{$q$-deformed Onsager symmetry in boundary integrable models\\ related to twisted U$_{q^{1/2}}(\widehat{sl_2})$ symmetry}\Author

\vspace{- 0.1mm}
 \Address

\vskip 0.6cm

\centerline{\bf Abstract}\vspace{0.3mm}
We consider a unified model, called ancestor model, associated with twisted trigonometric $R$ matrix
which model leads to several descendant integrable lattice models related to the U$_{q^{1/2}}(\widehat{sl_2})$ symmetry. 
Boundary operators compatible with integrability are introduced to this model. 
Reflection and dual reflection equations to ensure integrability of the system 
are shown to be same as the untwisted case.
It follows that underlying symmetry of the ancestor model with integrable boundaries 
is identified with the $q$-deformed analogue of Onsager's symmetry. 
The transfer matrix and its related mutually commuting quantities 
are expressed in terms of an abelian subalgebra in the $q$-Onsager algebra. 
It is illustrated that the generalized McCoy-Wu model with general open 
boundaries enjoys this symmetry.

\vspace{0.1cm} .
\vskip 0.8cm


\vskip -0.6cm


\section{Introduction} 
Unifying scheme to rich varieties of models has played important roles 
to develope our concepts and perspectives in all scientific fields. 
Especially in theoretical physics and mathematics, most of successes of such attempts 
is deeply connected with findings of underlying symmetries, which brings beauty and simplicity 
into relief. Although such symmetries may be hidden, in many cases they are codified by groups which are
generated by Lie algebra and its generalizations including the Kac-Moody algebra and quantum algebra. 
One of the most direct application in this subject is in the theory of 
nontrivial completely integrable models (continuum or lattice) in 1+1 space time dimensions.
Their models have an (in)finite set of sufficient mutually 
commuting conserved quantities that all physically relevant information 
can be extracted without any approximation in principle.
Therefore the interesting features of integrabilities 
have made profound influence between many branches of physics and mathematics,
for example (super)string theory and condensed matter systems in physics and
quantum Lie algebra, $q$-orthogonal polynomials, $q$-difference equations and knot theory in mathematics 
\cite{BPZ,Pol,DJF,Koo,RT}.

Since the seminal works by Cherednik \cite{Che} and Sklyanin \cite{Skl}, 
quantum boundary integrable models have received a lot of attention 
in the context of systematic studies on the XXZ anisotropic open spin chain models \cite{CLSW,Nepxxz,NRG,YZ07,Doik}, 
two dimensional boundary integrable quantum field theory \cite{GZ,Gho,BD,DM,BK03,FK,CDRS,Doik08}, 
the open string/spin chain sector of AdS/CFT correspondence in super string theory 
\cite{HM,AN,MN,Lin} and so on. In recent years, based on the Sklyanin's dressing method,
a breakthrough emerged \cite{Bas1,Bas2} from the study of the algebraic structure encoded in the reflection
equation for certain boundary conditions and U$_{q^{1/2}}(\widehat{sl_2})$ trigonometric $R$-matrix.
Furthermore successive studies in this line revealed that the transfer matrix of the model
(including the XXZ open spin chain model with general boundary conditions) is generated by an abelian subalgebra of the $q$-Onsager algebra \cite{BK0503,BK0507}. 
The $q$-Onsager algebra is also referred as the tridiagonal algebra, 
which is first introduced by Terwilliger in the context of the P and Q- association schemes related to 
the Askey scheme of orthogonal polynomials in the mathematical side \cite{Ter}. 
The $q$-Onsager algebra can be viewed 
as an extension of the $q$-Serre relations and a $q$-anlogue of the Dolan-Grady relations \cite{DolGra} that define the Onsager algebra \cite{Ons}.
It is worth emphasizing that the Onsager algebra not only have played crucial roles in planar Ising model, XY model and superintegrable chiral Potts model \cite{Ons,Davies,GehRit,Bax,ND,IPSTG,YaPerk}, but so much of what we understand in generality was contained in 
the Onsager's solution that led us to the discovery of the Yang-Baxter equation
(the star triangular relation) \cite{Per}. It seems that the emergenece of the $q$-Onsager algebra opens the new possibility 
of studying massive continum or lattice quantum integrable models.

New possibilities to deform quantum integrable (spin chain) models using the
Drinfeld twist \cite{Drin} have already been considered \cite{Kun,KSMND}.
We employ their ideas to construct much wider class of quantum integrable lattice systems and introduce integrable boundaries for the ancestor model \cite{Kun} related to the U$_q(\widehat{sl_2})$ case.
Then matrix elements of the associated quantum Lax operator 
is made of $t$ (or $\theta$)-deformed extension of the trigonometric Sklyanin algebra \cite{Kun},
which describe several quantum integrable models related to the twisted U$_q(\widehat{sl_2})$ in a unified way.
The main purpurse in the article is to identify underlying symmetry 
in the ancestor model with open integrable boundary conditions,
by applying the Sklyanin's dressing method with the Lax operator.

This article is organized in the following way:
in the section 2, a brief formulation of quantum inverse scattering method (QISM) 
for quantum integrable systems and its integrable boundary extension is summarized. 
The Lax operator for the twisted U$_q(\widehat{sl_2})$ case is uniquely determined except for the gauge degreee of freedom
up to the Laurant polynomials of degree 1 in the section 3. 
To utilize the Sklyanin's dressing method, we need to find the inverse representation of the Lax operator. 
To find the inverse operator can be performed by introducing a new operator. It is shown
 that a new operator 
unentangles nontriviality of the twisted transformation to the untwisted algebra.
In the section 4, it is shown that reflection algebra resulting from the reflection equation 
for the twisted U$_q(\widehat{sl_2})$ case is the same as the untwisted one (the extended trigonometric Sklyanin algebra).
As shown in Baseilhac and Shigechi \cite{BS}, it is clear that the $q$-Onsager symmetry 
appears in the reflection algebra. We explicitly construct the Sklyanin's dressed solution with the c-number solution by Goshal and Zamoldochikov.
As expected, the foundamental operators for $N=1$ dressed solution are satisfied with the Askey-Wsilson relations.   
Through the same procedure in the articles \cite{BK0503,BK0507}, general $N$ dressed solution is explicitly constructed.
The fundamental operators are satisfied with the tridiagonal relation, so that the transfer matrix enjoys the $q$-Onsager symmetry.
As described by linear combinations of generators of the $q$-Onsager symmetry, the transfer matrix is made from the abelian 
subalgebra of the $q$-Onsager algebra. As the application of this construction, we derived the Hamiltonian of  a generalized McCoy-Wu model \cite{MW,Kun} with general boundary conditions in the section 5. 
The model may be considered as a generalization of XXZ spin chain with Dzyyaloshinsky-Moriya interactions \cite{DzMo}.
All of the integral motions are identified as an abelian subalgebra of the $q$-Onsager algebra. The discussion is given in the section 6.

\section{Formulation on quantum integrable lattice systems}
This section serves as a brief introduction by quantum inverse scattering method \cite{TSKF}
for quantum lattice systems with periodic and open boundaries compatible with integrability.

Complete integrablity of quantum systems takes its roots on the existence of 
(in)finite set of mutually conserved quantities to be involutive. Within quantum 
inverse scattering method, this is encoded in the existence of monodromy matrix $T(u)$ 
satisfying the quantum Yang-Baxter equation(QYBE)
\beq
    R_{12}(u,v)\stackrel{1}{T}(u)\stackrel{2}{T}(v)
                    =\stackrel{2}{T}(v)\stackrel{1}{T}(u)R_{12}(u,v),
\eeq
where $\stackrel{1}{T}=T\otimes I, \stackrel{2}{T}=I\otimes T$, which equation is 
in auxiliary spaces $V_1\otimes V_2$ and a full Hilbert space ${\cal H}$ acted by the monodromy matrix $T$. 
Taking the trace of the above equation, we obtain
\beq
    {\rm Tr}_{12}(\stackrel{1}{T}(u)\stackrel{2}{T}(v))
               ={\rm Tr}_{12}(R^{-1}_{12}(u,v)\stackrel{2}{T}(v)\stackrel{1}{T}(u)R_{12}(u,v))
               ={\rm Tr}_{12}(\stackrel{2}{T}(v)\stackrel{1}{T}(u)),
\eeq
where ${\rm Tr}_{12}(\stackrel{1}{T}(u)\stackrel{2}{T}(v))$ defines the trace 
over $V_1\otimes V_2$. By using the definitive relation ${\rm Tr}_{12}(\stackrel{1}{T}(u)\stackrel{2}{T}(v))={\rm Tr}(T(u)){\rm Tr}(T(v))$, 
it is shown that the transfer matrix $t(u)={\rm Tr}(T(u))$ is satisfied with 
the trivial commutation relation
\beq
    [t(u),t(v)]=0,\quad({\rm for\ any\ variables}\ u\ {\rm and}\ v).
\eeq
The above equation guarantees that the transfer matrix poseeses mutually 
commuting quantities $ I_{2k+1}$ as the from $t(u)=\sum {\cal F}_{2k+1}(u) I_{2k+1}$. 
As the result, the integrability of the quantum systems is ensured in the quantum version of 
the Liouville sense defined in the context of the classical case.

For $N$-site periodic lattice models, the monodromy matrix $T(u)$ is discribed by 
\beq
    T(u)=\displaystyle L_N(u) L_{N-1}(u)\cdots L_2(u)L_1(u)
\eeq
in terms of the Lax operator $L_i(u)$ at the $i$-th lattice site 
acting on a local Hilbert space ${\cal H}_i$. Requiring ultralocality for the Lax operator, 
$[\stackrel{1}{L}_{i},\stackrel{2}{L}_{j}]=0\ (i\ne j)$, the QYBE leads to
\beq
    \label{qybeq}
     R_{12}(u,v)\stackrel{1}{L}_{i}(u)\stackrel{2}{L}_{i}(v)=\stackrel{2}{L}_{i}(v)\stackrel{1}{L}_{i}(u)R_{12}(u,v),\quad (i=1,\cdots,N)
\eeq
as a consequence of its local version. Supposed that the associativity for the triple product 
$\stackrel{1}{L}_{i}(u)\stackrel{2}{L}_{i}(v)\stackrel{3}{L}_{i}(w)$ holds, the equation in effect 
turns into an independent equation for the matrix $R(u)$ known as the Yang-Baxter equation (YBE)
\beq
\label{YBE}
    R_{12}(u,v)R_{13}(u,w)R_{23}(v,w)=R_{23}(v,w)R_{13}(u,w)R_{12}(u,v).
\eeq
For ultralocal lattice models that we are intrested in, the matrix $R_{12}(u,v)$ takes 
the form of $R_{12}(uv^{-1})$, ${\it i.e.}$ the matrix elements of $R$ are given through 
functions of the spectral parameters $uv^{-1}$. 

For quantum integrable systems with open boundaries, 
each set of boundary conditions is associated with a choice of boundary operators $K_{\pm}(u)$.
Without spoiling the nice algebraic structure and analytic properties of bulk integrability,
the adequete conditions for the  boundary operators $K_{\pm}(u)$ compatiable with integrability, respectively, 
are obtained from the following equations known as the reflection and  dual reflection equations \cite{Che,Skl,MNKSNADR}:
\beqa
\label{reflect}
    &&R_{12}(uv^{-1})(K_-(u)\otimes 1) R_{12}^{t_1t_2}(uv)(1\otimes K_-(v))
        =(1\otimes K_-(v))R_{12}(uv)(K_-(u)\otimes 1) R_{12}^{t_1t_2}(uv^{-1}),\\
   \noalign{\vskip 2mm}
\label{dreflect}
    &&R_{12}(u^{-1}v)(K^{t_1}_+(u)\otimes 1)(M^{-1}\otimes 1) R_{12}^{t_1t_2}(q^{-1}u^{-1}v^{-1})(M\otimes 1)(1\otimes K^{t_2}_+(v))\non\\
    &&\hspace{3cm}=(1\otimes K^{t_2}_+(v))(M\otimes 1)R_{12}(q^{-1}u^{-1}v^{-1})(M^{-1}\otimes 1)(K^{t_1}_-(u)\otimes 1) R_{12}^{t_1t_2}(u^{-1}v),
\eeqa
where the symbol $t_i$ denotes transposition in the $i$-th auxiliary space. Furthermore the matrix $M$ is determined by the following relation \cite{RSTS}
\beq
 \label{Mmatrix}
    \{\{\{R_{12}^{t_2}(u)\}^{-1}\}^{t_2}\}^{-1}
         =\frac{\zeta(q^{1/2}u)}{\zeta(qu)}(1\otimes M)R_{12}(qu)(1\otimes M)^{-1},
    \quad M^t=M,
\eeq
where $\zeta(u)I\!\!I=R_{12}(u)R_{21}(u^{-1})$ with the unit matrix $I\!\!I$.
Once finding out a solution of the reflection equation (\ref{reflect}) for $K_-(u)$, one can verify that
$K_+(u)$ matrix defined by
\beq
    K_+(u)=K_-^t(q^{-1/2}u^{-1})M
\eeq
satisfies the dual reflection equation (\ref{dreflect}). 
The transfer matrix for the $N$-site lattice model with the integrable open boundary conditions is defined by
\beq
    t(u)={\rm Tr}(K_+(u)K_-^{(N)}(u))),
\eeq
where
\beq
    K_-^{(N)}(u)=\left(\displaystyle\left(L_N(uv_N)\cdots L_1(uv_1)\right)K_-(u)
    \left(L_1^{-1}(u^{-1}v_1)\cdots L_N^{-1}(u^{-1}v_N)\right)\right).
\eeq
Interestingly the dressed boundary operator $K_-^{(N)}(u)$ is also a solution of the reflection equation. 
This permits us to make a sequence of the operator-valued solutions from one solution of the reflection equation. 
This construction is known as the Sklyanin's dressing method. 
Thanks to these algebraic relations, the transfer matrix commutes for any spectral parameters $u$ and $v$:
\beq
    [t(u),t(v)]=0,
\eeq
which is enough to ensure the integrabiity of the system even with open boundaries.

To construct much wider class of integrable models starting from a solution of YBE for the $R$-matrix, 
let's look into a transformation called twist introduced in the article\cite{Drin}. 
Provided that the twisting operator ${\cal F}$ satisfies the conditions
\beq
    R_{12}(u){\cal F}_{12}{\cal F}_{23}={\cal F}_{23}{\cal F}_{13}R_{12}(u),\qquad
    {\cal F}_{12}{\cal F}_{13}{\cal F}_{23}={\cal F}_{23}{\cal F}_{13}{\cal F}_{12},\qquad
    {\cal F}_{ij}={\cal F}_{ji}^{-1},
\eeq
we obtain the twisted $R$-matrix
\beq
    {\tilde R}_{12}(u)={\cal F}_{12}^{-1}R_{12}(u){\cal F}_{12}^{-1}
\eeq
from the original one $R_{12}(u)$. Corresponding to the transformation, 
twisted Lax operator and twisted boundary operators are obtained.

\section{Lax operator related to twisted trigonometric $R$-matrix}

The Yang-Baxter equation (\ref{YBE}) restricts the solution of the $R$-matrix to integrable models.
The most simplest solution of YBE acting on the two dimensional auxiliary spaces $V_1\otimes V_2$ is 
the trigonometric solution related to U$_{q^{1/2}}$($\widehat{sl_2}$) symmetry:
\beq
\label{trigonormatrix}
    R_{12}(u)=\left(\begin{array}{cccc}
                            a(u) &        &        &    \\
                                 &b(u)    &\tilde c&    \\
                                 &\tilde c& b(u)   &    \\
                                 &        &        &a(u)\\
                       \\\end{array}\right),
\eeq
where $a(u)=q^{1/2}u-q^{-1/2}u^{-1},\ b(u)=u-u^{-1},\ \t c=q^{1/2}-q^{-1/2}$. 
Starting from the trigonometric solution (\ref{trigonormatrix}) of YBE, 
one can produce the twisted trigonometric $R$-matrix  
in terms of a suitable representation ${\cal F}_{12}=e^{i\theta(\sigma_3/2\otimes Z-Z\otimes\sigma_3/2)}$ 
as the twisted operator, where $Z$ is the central charge. As a result, we obtain
\beq
    {\tilde R}_{12}(u)=\left(\begin{array}{cccc}
                            a(u) &        &        &    \\
                                 &tb(u)    &\tilde c&    \\
                                 &\tilde c& t^{-1}b(u)   &    \\
                                 &        &        &a(u)\\
                       \\\end{array}\right),
\eeq
where $t=e^{-2i\theta Z}$. The twisting procedure enables the Lax operator to take the different twist parameter $t_n$ at each lattice site. Then the related twisted Lax operator $L_n(u)$ at the $n$-th lattice site takes 2$\times$2 matrix form in the auxialiary space as follows:
\begin{equation}
\label{originallax}
    L_n(u)=\left(\begin{array}{cc}
                u\tau_1^{-}+u^{-1}\tau_{1}^{+}&\tau_{12}                     \\
                \tau_{21}                     &u\tau_2^{-}+u^{-1}\tau_{2}^{+}\\
               \end{array}\right).
\end{equation}
These operators $\tau_{i}^{\pm}$ and $\tau_{ij}$ ($i,j=1,2$) satisfy the 
$t$-deformation of the extended trigonometric Sklyanin algebra \cite{Kun}:
\begin{eqnarray}
\label{tdefETSA}
    &&[\tau_i^{\pm},\tau_{j}^{\pm}]= [\tau_i^{\pm},\tau_{j}^{\mp}]=0,\non\quad\\
  \noalign{\vskip 2mm}
    &&\tau_{1}^{\pm}\tau_{12}=t_n^{-1}q^{\pm1/2}\tau_{12}\tau_{1}^{\pm},\quad
    \tau_{2}^{\pm}\tau_{12}=t_n^{-1}q^{\mp1/2}\tau_{12}\tau_{2}^{\pm},\non\\
  \noalign{\vskip 2mm}
    &&\tau_{1}^{\pm}\tau_{21}=t_n q^{\mp1/2}\tau_{21}\tau_{1}^{\pm},\quad
    \tau_{2}^{\pm}\tau_{12}=t_n q^{\pm1/2}\tau_{21}\tau_{2}^{\pm},\quad \non\\   
  \noalign{\vskip 2mm}
    &&t_n\tau_{21}\tau_{12}-t_n^{-1}\tau_{12}\tau_{21}={\tilde c}(\tau_{1}^+\tau_{2}^- -\tau_{1}^{-}\tau_{2}^+).
\end{eqnarray}
The coproduct structure for this algebra is found by the elegant formulation by Faddeev-Reshetikhin-Takhtajan \cite{FRT}. 
Symmetric and nonsymmetric realizations of the algebra can generate several descendant lattice models 
without limiting procedures, for examples the generalized McCoy-Wu model and 
$t$-deformation of such models as the Liouville lattice model, $q$-oscillator model related to the $sl_q(2)$ and $sl_q(1,1)$ \cite{Kundu07}, 
the derivative non-linear Schrodinger equation, lattice version of the massive Thirring model, lattice sine-Gordon and  
their hybrid models \cite{Kundu06} and so on. 
Although one could explicitly construct the transfer matrix for these integrable models with periodic boundary conditions, 
the derivation is out of our intrests in the present aricle.

Our main aim in this article is to introduce intgrable open boundaries 
for these models and construct the transfer matrix and mutually commuting quantities in terms of generators in the $q$-Onsager symmetry. 
For our purpurse, it is useful to factorize the above opearators (\ref{tdefETSA}) in terms of new operators
\beq
    \tau_{i}^\pm=\t\tau_{i}^{\pm}\tau_g,
    \tau_{12}=t_n^{-1/2}\t\tau_{12}\tau_g,\quad
    \tau_{21}=t_n^{1/2}\t\tau_{21}\tau_g,
\eeq
with the following algebraic relations
\beq
    \tau_g\t\tau_{12}=t_n^{-1}\t\tau_{12}\tau_g,\quad
    \tau_g\t\tau_{21}=t_n\t\tau_{21}\tau_g,\quad
    [\tau_g,\t\tau_{i}^{\pm}]=0.
\eeq
Plugging the factorization of the operators into Eq.(\ref{tdefETSA}), they reduce to the algebraic relations
on the non-twisted extended trigonometric Sklyanin algebra:
\beqa
\label{unETSA}
    &&\t\tau_{1}\tilde\tau_{12}= q^{\pm1/2}\t\tau_{12}\t\tau_{1}^{\pm},\quad
      \t\tau_{2}\tilde\tau_{12}= q^{\mp1/2}\t\tau_{12}\t\tau_{2}^{\pm},\quad\non\\
  \noalign{\vskip 2mm}
    &&\t\tau_{1}\tilde\tau_{21}= q^{\mp1/2}\t\tau_{21}\t\tau_{1}^{\pm},\quad
      \t\tau_{2}\tilde\tau_{21}= q^{\pm1/2}\t\tau_{21}\t\tau_{2}^{\pm},\quad\non\\
  \noalign{\vskip 2mm}
    &&[\t\tau_{21},\t\tau_{12}]={\t c}(\t\tau_1^{+}\t\tau_{2}^{-}-\t\tau_{1}^{-}\tau_{2}^{+}),
\eeqa
It turns out that the introduction of these new operators unpicks 
the entanglement of twisted quadratic algebra. 
As usual, it is shown that there exist five Casimir operators $w_\pm, w_{0_1}, w_{0_2}, w$: 
\beqa
\label{Casimir}
   &&\hspace{2cm}\t\tau_1^{\pm}\t\tau_{2}^{\pm}=w_{\pm},\quad  
   \t\tau_i^{-}\t\tau_{i}^{+}=w_{0_i},\quad(i=1,2),\non\\
   &&\t\tau_{12}\t\tau_{21}-q^{1/2}\t\tau_{1}^{-}\t\tau_{2}^{+}-q^{-1/2}\t\tau_{1}^{+}\t\tau_{2}^{-}
   =\t\tau_{21}\t\tau_{12}-q^{-1/2}\t\tau_{1}^{-}\t\tau_{2}^{+}-q^{1/2}\t\tau_{1}^{+}\t\tau_{2}^{-}
   =w,
\eeqa
with the relation $w_-w_+=w_{0_1}w_{0_2}$.

The Lax operator (\ref{originallax}) is re-expressed in terms of these algebraic elements:
\beq
    L_n(u)=\left(\begin{array}{cc}
                   u\t\tau_{1}^{-}+u^{-1}\t\tau_{1}^+& t_n^{-1/2}\t\tau_{12}               \\
                   t_n^{1/2}\t\tau_{21}                & u\t\tau_{2}^{-}+u^{-1}\t\tau_{2}^+\\
               \end{array}\right)\tau_g.
\eeq
As these consequences, we obtain the operator $\t L_n(u)$ 
\beq
    \t L_n(u)=\tau_g^{-1}
            \left(\begin{array}{cc}
                   -(q^{-1/2}u\t\tau_{2}^{-}+q^{1/2}u^{-1}\t\tau_{2}^+)& t_n^{-1/2}\t\tau_{12}               \\
                   t_n^{1/2}\t\tau_{21}                & -(q^{-1/2}u\t\tau_{1}^{-}+q^{1/2}u^{-1}\t\tau_{1}^+)\\
                  \end{array}\right),
\eeq
which is proportional to the inverse of the Lax operator. In fact, it is easy to check
\beq
    L_n(u)\t L_n(u)=\rho(u)I\!\!I, \quad \rho(u)=w-(q^{-1/2}w_-u^2+q^{1/2}w_+u^{-2})
\eeq
with the algebraic relations (\ref{unETSA}) and the Casimir operators (\ref{Casimir}).

\section{Reflection equation and dressing $K$ matrix related to twisted $R$-matrix}
In the previous section we derived the fundamental parts to construct dressed $K$-matrix,
although solutions for the reflection equation and the dual reflection equation are still left. 
We explicitly construct thier solutions in this section.

Before finding solutions of the reflection equation and dual reflection equation 
for boundary operators $K_\pm(u)$, it is worthwhile writing out all elements of the reflection equation. 
Supposed that
\[
    K{_-}(u)=\left(\begin{array}{cc}
                   A(u)&B(u)\\
                   C(u)&D(u)
                \end{array}
          \right),
\]
the reflection equation for $K_-$ reduces to the sixteen algebraic equations: 
\beqa
\label{16Kmeq}
    &&(1)\ a_-\t c(BC^\prime-B^\prime C)+a_-a_+[A,A^\prime]=0,\non\\
 \noalign{\vskip 2mm}
    &&(2)\ a_-\t c(CB^\prime-C^\prime B)+a_-a_+[D,D^\prime]=0,\non\\
 \noalign{\vskip 2mm}
    &&(3)\ b_-b_+[A,D^\prime]+\t c^2[D,D^\prime]+\t c a_+(CB^\prime-C^\prime B)=0,\non\\
 \noalign{\vskip 2mm}
    &&(4)\ b_-b_+[D,A^\prime]+\t c^2[A,A^\prime]+\t c a_+(BC^\prime-B^\prime C)=0,\non\\
 \noalign{\vskip 2mm}
    &&(5)\ \t cb_+(DA^\prime-D^\prime A)+b_-\t c(AA^\prime-DD^\prime)+b_-a_+[B,C^\prime]=0,\non\\
 \noalign{\vskip 2mm}
    &&(6)\ \t cb_+(AD^\prime-A^\prime D)+b_-\t c(DD^\prime-AA^\prime)+b_-a_+[C,B^\prime]=0,\non\\
 \noalign{\vskip 2mm}
    &&(7)\ b_-b_+AC^\prime+\t c^2 DC^\prime+\t c a_+ CA^\prime- a_-a_+C^\prime A-a_-\t c D^\prime C=0\non\\
 \noalign{\vskip 2mm}
    &&(8)\ b_-b_+DB^\prime+\t c^2 AB^\prime+\t c a_+ BD^\prime- a_-a_+B^\prime D-a_-\t c A^\prime B=0,\non\\
 \noalign{\vskip 2mm}
    &&(9)\  b_-b_+B^\prime A+\t c^2 B^\prime D+\t c a_+ A^\prime B- a_-a_+AB^\prime -a_-\t c BD^\prime =0,\non\\
 \noalign{\vskip 2mm}
    &&(10)\ b_-b_+C^\prime D+\t c^2 C^\prime A+\t c a_+ D^\prime C- a_-a_+DC^\prime -a_-\t c CA^\prime =0,\non\\
 \noalign{\vskip 2mm}
    &&(11)\ b_-a_+BD^\prime+\t cb_+DB^\prime+b_-\t cAB^\prime-a_-b_+D^\prime B=0,\non\\
 \noalign{\vskip 2mm}
    &&(12)\ b_-a_+CA^\prime+\t cb_+AC^\prime+b_-\t cDC^\prime-a_-b_+A^\prime C=0,\non\\
 \noalign{\vskip 2mm}
    &&(13)\ b_-a_+A^\prime B+\t cb_+B^\prime A+b_-\t cB^\prime D-a_-b_+B^\prime A=0,\non\\
 \noalign{\vskip 2mm}
    &&(14)\ b_-a_+D^\prime C+\t cb_+C^\prime D+b_-\t cC^\prime A-a_-b_+C^\prime D=0,\non\\
 \noalign{\vskip 2mm}
    &&(15)\ a_-b_+[B,B^\prime]=0,\non\\
 \noalign{\vskip 2mm}
    &&(16)\ a_-b_+[C,C^\prime]=0,
\eeqa
where we used the notations $a_-=a(u/v),\ a_+=a(uv)$ and similarly for $b$.
Also $A=A(u), A^\prime=A(v)$ and similarly for $B,C$ and $D$. 
Instead of writing these equations for $K_+$, we solve the equation (\ref{Mmatrix}). 
The matrix $M$ is determined as the unit matrix
\beq
    M=\left(\begin{array}{cc}1&0\\0&1\end{array}\right).
\eeq 
Therefore solutions for the boundary operator $K_+$ of the dual reflection equation is easily deriven.
It is interesing to note that these sixteen algebraic equations for $K_-(u)$ do not depend on the twisted parameter $t$, 
{\it i.e.} there is no difference from the untwisted case to the twisted one related to U$_q(\widehat{sl_2})$. 
Therefore the class of solution for Eq.(\ref{16Kmeq}) belongs to one analysed by Baseilhac and Shigechi \cite{BS}.

The most simplest solution for the above equations consists of $c$-number elements, which is given by
\beq
    K_{-}^{c}(u)=\left(\begin{array}{cc}
                      u\epsilon_++u^{-1}\epsilon_{-}&\frac{k_+}{\t c}(u^2-u^{-2})\\
                      \frac{k_-}{\t c}(u^2-u^{-2})&u\epsilon_-+u^{-1}\epsilon_+\\
                   \end{array}\right),
\eeq
where $\epsilon_{\pm}$ and $k_{\pm}$ are free parameters of the theory \cite{GZ,DG}.
Similarily, $c$-number solution of the dual reflection equation is
\beq
    K_{+}^c(u)=\left(\begin{array}{cc}
                      q^{1/2}u\kappa+q^{-1/2}u^{-1}\kappa^\ast&\kappa_+(q^{1/2}+q^{-1/2})(qu^2-q^{-1}u^{-2})\\
                      \kappa_-(q^{1/2}+q^{-1/2})(qu^2-q^{-1}u^{-2})&q^{1/2}u\kappa^\ast+q^{-1/2}u^{-1}\kappa\\
                   \end{array}\right),
\eeq
which parameters $\kappa,\kappa^\ast$ and $\kappa_{\pm}$ are also free parameters.

\subsection{$N=1$ dressed solution and Askey-Wilson relation}
To construct general $N$-dressing $K$-matrix, let us start from deriving the $N=1$ dressed solution 
for the reflection algebra. By using the operator $\t L(u)$ instead of the inverse operator of $L(u)$, 
the $N=1$ dressed solution 
\beq
    K^{(1)}_-(u)=L_1(uv_1)K_-^{c}(u)\t L_{1}(u^{-1}v_1)
              =\left(\begin{array}{cc}
                      {\cal A}^{(1)}(u)&{\cal B}^{(1)}(u)\\
                      {\cal C}^{(1)}(u)&{\cal D}^{(1)}(u)\\
                     \end{array}\right)
\eeq
is obtained as the following forms 
\beqa
    &&{\cal A}^{(1)}(u)=u\epsilon_{+}^{(1)}+u^{-1}\epsilon_{-}^{(1)}
                +(u^2-u^{-2})\left(q^{1/2}uW_{0}^{(1)}-q^{-1/2}u^{-1}W_{1}^{(1)}\right),\non\\
  \noalign{\vskip 2mm}
    &&{\cal D}^{(1)}(u)=u\epsilon_{-}^{(1)}+u^{-1}\epsilon_{+}^{(1)}
                +(u^2-u^{-2})\left(q^{1/2}uW_{1}^{(1)}-q^{-1/2}u^{-1}W_{0}^{(1)}\right),\non\\
  \noalign{\vskip 2mm}
    &&{\cal B}^{(1)}(u)=-\frac{(u^2-u^{-2})}{k_-w^{(1)}_{0_2}}
                      \left(\frac{k_+k_-w^{(1)}_-w^{(1)}_+(q^{1/2}u^2+q^{-1/2}u^{-2})}{\t c}
                                    +\frac{G_1^{(1)}}{q^{1/2}+q^{-1/2}}+\om_0^{(1)}\right),\non\\
  \noalign{\vskip 2mm}
    &&{\cal C}^{(1)}(u)=-\frac{(u^2-u^{-2})}{k_+w^{(1)}_{0_1}}
                      \left(\frac{k_+k_-w^{(1)}_-w^{(1)}_+(q^{1/2}u^2+q^{-1/2}u^{-2})}{\t c}
                                    +\frac{\t G_1^{(1)}}{q^{1/2}+q^{-1/2}}+\om_0^{(1)}\right), 
\eeqa
where $w_\pm^{(j)}$ and $w^{(j)}$ represents the Casimir operators (\ref{Casimir}) of the extended Sklyanin algebra (\ref{unETSA}) for $j$-th Lax operators. The parameters $\epsilon_{\pm}^{(1)}$ and $\omega_{0}^{(1)}$ are given by
\beqa
    &&\epsilon_{\pm}^{(1)}=-(w_-^{(1)}q^{-1/2}v_1^2+w_+^{(1)}q^{1/2}v_1^{-2})\epsilon_{\pm}
                                        +w^{(1)}\epsilon_{\mp},\quad\non\\
    &&\omega_0^{(1)}=-\frac{k_+k_-w^{(1)}(w_-^{(1)}q^{-1/2}v_1^2+w_+^{(1)}q^{1/2}v_1^{-2})+\t c^2\epsilon_+\epsilon_-w_-^{(1)}w_+^{(1)}}{q-q^{-1}}\non.
\eeqa
Similarily in \cite{BK0503,BK0507}, the generators $G_1^{(1)}$ and $\t G_{1}^{(1)}$ are given by 
$G_1^{(1)}=[W_{1}^{(1)},W_{0}^{(1)}]_q$ and $\t G_1^{(1)}=[W_{0}^{(1)},W_{1}^{(1)}]_q$ in terms of the $q$-commutator
\beq
    [X,Y]_q=q^{1/2}XY-q^{-1/2}YX.
\eeq
Explicit representations of the generators $W_0^{(1)},W_{1}^{(1)},G_{1}^{(1)},\t G_{1}^{(1)}$ are written by 
\beqa
    && W_0^{(1)}=\frac{1}{\t c}t_1^{1/2}{k}_+v_1{\t \tau}_{21}{\t \tau_1}^-
                -\frac{1}{\t c}t_1^{-1/2}{k}_-v_1^{-1}{\t \tau}_{12}{\t \tau_2}^+
                -\epsilon_+{\t \tau_1}^-{\t \tau_2}^+ ,\non\\
  \noalign{\vskip 2mm}
    && W_1^{(1)}=-\frac{1}{\t c}t_1^{1/2}{k}_+v_1^{-1}{\t \tau}_{21}{\t \tau_1}^+
                +\frac{1}{\t c}t_{1}^{-1/2}{k}_-v_1{\t \tau}_{12}{\t \tau_2}^-
                -\epsilon_-{\t \tau_1}^+{\t \tau_2}^-\non\\
  \noalign{\vskip 2mm}
    && G_1^{(1)}=\frac{-w^{(1)}_{0_2}t_1^{-1}(q^{1/2}+q^{-1/2})k_-^2}{\t c}\t\tau_{12}^2
                    +\frac{w_{0_2}^{(1)}(q^{1/2}+q^{-1/2})k_+k_-}{\t c}
                             (q^{-1/2}v_1^2(\t\tau^-_1)^2+q^{1/2}v_1^{-2}(\t\tau_1^+)^2)
                                  +\t cw_{0_1}^{(1)}w_{0_2}^{(1)}\epsilon_-\epsilon_+\non\\
    && \hspace{0.5cm} 
                    -w_{0_2}^{(1)}k_-t_1^{-1/2}(q^{1/2}+q^{-1/2})
                      (-q^{1/2}v_1^{-1}\epsilon_-\t\tau_{12}\t\tau_{1}^+
                                  +q^{-1/2}v_1\epsilon_+\t\tau_{12}\t\tau_{1}^-)
                      +\frac{k_+k_-w^{(1)}(q^{-1/2}w_-^{(1)}v_1^2+q^{1/2}w_+^{(1)}v_1^{-2})}{w_{0_1}^{(1)}w_{0_2}^{(1)}\t c},\non\\
   \noalign{\vskip 2mm}
    &&\t G_1^{(1)}=\frac{-w^{(1)}_{0_1}t_1(q^{1/2}+q^{-1/2})k_+^2}{\t c}\t\tau_{21}^2
                    +\frac{w_{0_1}^{(1)}(q^{1/2}+q^{-1/2})k_+k_-}{\t c}
                             (q^{-1/2}v_1^2(\t\tau^-_2)^2+q^{1/2}v_1^{-2}(\t\tau_2^+)^2)
                                  +\t cw_{0_1}^{(1)}w_{0_2}^{(1)}\epsilon_-\epsilon_+\non\\
    && \hspace{0.5cm} 
                    -w_{0_1}^{(1)}k_+t_1^{1/2}(q^{1/2}+q^{-1/2})
                      (-q^{1/2}v_1^{-1}\epsilon_+\t\tau_{21}\t\tau_{2}^+
                                  +q^{-1/2}v_1\epsilon_-\t\tau_{21}\t\tau_{2}^-)
                      +\frac{k_+k_-w^{(1)}(q^{-1/2}w_-^{(1)}v_1^2+q^{1/2}w_+^{(1)}v_1^{-2})}{w_{0_1}^{(1)}w_{0_2}^{(1)}\t c}.\non
\eeqa

Straightforward calculations show that the fundamental generators $W_{0}^{(1)}$ and $W_{1}^{(1)}$ satisfy the Askey-Wilson relations \cite{Zhe}
\beqa
    &&[W_1^{(1)},[W_1^{(1)},W_0^{(1)}]_q]_{q^{-1}}=(q^{1/2}+q^{-1/2})^2k_-k_+w_{-}^{(1)}w_{+}^{(1)}W_{0}^{(1)}\non\\
    &&                           \hspace{5cm}+(q-q^{-1})\om_0^{(1)}W_{1}^{(1)}-(q^{1/2}+q^{-1/2})k_+k_-w_-^{(1)}w_+^{(1)}\epsilon_-^{(1)},\non\\
  \noalign{\vskip 2mm}
    &&[W_0^{(1)},[W_0^{(1)},W_1^{(1)}]_q]_{q^{-1}}=(q^{1/2}+q^{-1/2})^2k_-k_+w_{-}^{(1)}w_{+}^{(1)}W_{1}^{(1)}\non\\
    &&                           \hspace{5cm}+(q-q^{-1})\om_0^{(1)}W_{0}^{(1)}-(q^{1/2}+q^{-1/2})k_+k_-w_-^{(1)}w_+^{(1)}\epsilon_+^{(1)},
\eeqa
which lead to the $q$-Dolan Grady relations
\beq
    [W_1^{(1)},[W_1^{(1)},[W_1^{(1)},W_0^{(1)}]_q]_{q^{-1}}]=\rho^{(1)}_0[W_1^{(1)},W_{0}^{(1)}],\quad
    [W_0^{(1)},[W_0^{(1)},[W_0^{(1)},W_1^{(1)}]_q]_{q^{-1}}]=\rho^{(1)}_1[W_0^{(1)},W_{1}^{(1)}],
\eeq
with $\rho^{(1)}_0=\rho^{(1)}_1=(q^{1/2}+q^{-1/2})^2k_-k_+w_{-}^{(1)}w_{+}^{(1)}$. For generic $q$, 
the generators $W_0^{(1)}, W_1^{(1)}$ are connected with the $q$-Racah polynomial and 
some related polynomials of the Askey scheme.

The transfer matrix for the $N=1$ deressed solution can be expressed in terms of these four generators $W_0^{(1)}, W_1^{(1)}, G_{1}^{(1)}, \t G_{1}^{(1)}$: 
\beqa
    t^{(1)}(u)&\!\!\!=\!\!\!& {\cal F}_0^{(1)}+(u^2-u^{-2})(qu^2-q^{-2}u^{-2}){\cal I}_1^{(1)},
\eeqa
where 
\beqa
    &&{\cal F}_0^{(1)}=(q^{1/2}+q^{-1/2})(\kappa^\ast \epsilon^{(1)}_+ +\kappa \epsilon^{(1)}_-)
                 +(q^{1/2}u^2+q^{-1/2}u^{-2})(\kappa \epsilon^{(1)}_+ +\kappa^\ast \epsilon^{(1)}_-)\non\\
              &&\hspace{5mm}-(q^{1/2}+q^{-1/2})(u^2-u^{-2})(qu^2-q^{-1}u^{-2})
                     \left( \frac{k_+k_-w_{-}^{(1)}w_{+}^{(1)}}{\t c}(q^{1/2}u^2+q^{-1/2}u^{-2})          
                                      +\omega^{(1)}_0\right)
                                     \left(\frac{\kappa_+}{k_+w^{(1)}_{0_1}}
                               +\frac{\kappa_-}{k_-w^{(1)}_{0_2}}\right)\nonumber
\eeqa
and
\beq
      {\cal I}_1^{(1)}=\left(\kappa W_{0}^{(1)}+\kappa^\ast W_{1}^{(1)}
                             -\frac{\kappa_+}{k_+w^{(1)}_{0_1}} \t G_{1}^{(1)}
                               -\frac{\kappa_-}{k_-w^{(1)}_{0_2}} G_{1}^{(1)}  \right).
\eeq
The algebraic part ${\cal I}_1$ does not depend on the spectral parameter $u$, so that 
it is expilicitly separeated from the functional one.
Then the eigenvalue problem of the transfer matrix is read in a different way as 
its problem of ${\cal I}_1$. Realizing the algebra (\ref{unETSA}) by the difference operators,
the eigenvalue problem of the conserved charge ${\cal I}_1$ leads to the second order difference equation.  
One can observe that eigenfunctions of ${\cal I}_1$ are associated with 
the $q$-hypergeometric function. The roots of polynomials, which obey the Bethe-Ansatz equation, 
is used to determine the spectrum of the system.
We do not penetrate further in the details. Instead, we recommend you to read the references \cite{WZ,Bas}.

\subsection{$N$-dressed solution}
To construct general $N$-site lattice models, we would like to find explict representation of $N$-dressed solution of the reflection equation.
After the above caluculation for the $N=1$ dressing method, the same kind of calucuration is performed for the $N=2$ case which result we suppress here.
Based on these results and our previous results \cite{BK0503,BK0507}, we can obatain the following form of the $N$-dressed solution
\beqa
\label{Ndressedsol}
    K^{(N)}_-(u)=L(uv_{N})\cdots L(uv_1) K^{c}_-(u) \t L(u^{-1}v_1)\cdots \t L(u^{-1}v_{N})
    =\left(
    \begin{array}{cc}
    {\cal A}^{(N)}(u)&{\cal B}^{(N)}(u)\\
    {\cal C}^{(N)}(u)&{\cal D}^{(N)}(u)
    \end{array}
    \right),
\eeqa
where these matrix elements are
\beqa
    &&{\cal A}^{(N)}(u)=u\epsilon_+^{(N)}+u^{-1}\epsilon_-^{(N)}
    +(u^2-u^{-2})\left(uq^{1/2}\sum_{k=0}^{N-1}P_{-k}^{(N)}W_{-k}^{(N)}
      -u^{-1}q^{-1/2}\sum_{k=0}^{N-1}P_{-k}^{(N)}W_{k+1}^{(N)}\right),\non\\
    &&{\cal D}^{(N)}(u)=u\epsilon_-^{(N)}+u^{-1}\epsilon_+^{(N)}
    +(u^2-u^{-2})\left(uq^{1/2}\sum_{k=0}^{N-1}P_{-k}^{(N)}W_{k+1}^{(N)}
      -u^{-1}q^{-1/2}\sum_{k=0}^{N-1}P_{-k}^{(N)}W_{-k}^{(N)}\right),\non\\
    &&{\cal B}^{(N)}(u)=\frac{(u^2-u^{-2})}{\displaystyle k_-\prod_{k=1}^{N}(-w_{0_2}^{(k)})}
    \left({\cal J}^{(N)}(u)+\frac{1}{q^{1/2}+q^{-1/2}}\sum_{k=0}^{N-1}P_{-k}^{(N)}(u)G_{k+1}^{(N)}\right),\non\\
    &&{\cal C}^{(N)}(u)=\frac{(u^2-u^{-2})}{\displaystyle k_+\prod_{k=1}^{N}(-w_{0_1}^{(k)})}
    \left({\cal J}^{(N)}(u)+\frac{1}{q^{1/2}+q^{-1/2}}\sum_{k=0}^{N-1}P_{-k}^{(N)}(u){\tilde G}_{k+1}^{(N)}\right).
\eeqa
Here the non algebraic parts $P_{-k}^{(N)},\ \epsilon_{\pm}^{(N)},{\cal J}^{(N)}(u)$ and $\omega_0^{(N)}$ are given by 
the following equations:
\beqa
    &&P_{-k}^{(N)}=-\frac{1}{q^{1/2}+q^{-1/2}}\sum_{n=k}^{N-1}
    \left(\frac{q^{1/2}u^2+q^{-1/2}u^{-2}}{q^{1/2}+q^{-1/2}}\right)^{n-k}
    C_n^{(N)}\non\\
    &&{\cal J}^{(N)}(u)=
               \frac{k_+k_-\displaystyle\prod_{k=1}^{N}(-w_{0_1}^{(k)})(-w_{0_2}^{(k)})}{q^{1/2}-q^{-1/2}}
               (q^{1/2}u^2+q^{-1/2}u^{-2})P_0^{(N)}(u)+\omega_0^{(N)}\non\\
    &&\epsilon_\pm^{(N)}=w^{(N)}\epsilon_{\mp}^{(N-1)}
    -(q^{-1/2}w_-^{(N)}v_N^2+q^{1/2}w_+^{(N)}v_N^{-2})\epsilon_{\pm}^{(N-1)}\non\\
    &&\omega_0^{(N)}=(-1)^N\frac{k_+k_-}{(q^{1/2}-q^{-1/2})}\prod_{k=1}^{N}\alpha_k(-w_{0_1}^{(k)})(-w_{0_2}^{(k)}),\non
\eeqa
with 
\beqa
    &&C_{-n}^{(N)}=(-1)^{N-n}(q^{1/2}+q^{-1/2})
    \sum_{k_1<\cdots<k_{N-n-1}=1}^{N}\alpha_{k_1}\cdots\alpha_{k_{N-n-1}}\non\\
    &&\alpha_1=\frac{q^{-1/2}w_-^{(1)}v_1^2+q^{1/2}w_+^{(1)}v_1^{-2}}{(q^{1/2}+q^{-1/2})w_{0_1}^{(1)}w_{0_2}^{(1)}}w^{(1)}
    +\frac{\epsilon_+^{(0)}\epsilon_-^{(0)}(q^{1/2}-q^{-1/2})^2}{k_+k_-(q^{1/2}+q^{-1/2})}\non\\
    &&\alpha_k=\frac{q^{-1/2}w_-^{(k)}v_k^2+q^{1/2}w_+^{(k)}v_k^{-2}}{(q^{1/2}+q^{-1/2})w_{0_1}^{(k)}w_{0_2}^{(k)}}w^{(k)}\non.
\eeqa
Reccursive representation of the algebraic parts $W_{k+1}^{(N)},\ W_{-k}^{(N)},\ G_{k+1}^{(N)},\ \tilde G_{k+1}^{(N)}$ is given by appendix A.
The form of the boundary K matrix (\ref{Ndressedsol}) is directly proved by mathematical induction 
dressing from the $N-1$ dressed solution to the $N$ dressed solution. Then as the consitency condition we find
the generalized linear combinations
\beqa
&&-\frac{(q^{1/2}-q^{-1/2})\omega_0^{(N)}}{k_+k_-\displaystyle\prod_{k=1}^{N}(-w_{0_1}^{(k)})(-w_{0_2}^{(k)})}W_{-l}^{(N)}+\sum_{k=1}^{N}C_{-k+1}^{(N)}W_{-k-l}^{(N)}+\epsilon_{(-)^l}^{(N)}=0\non\\
&&-\frac{(q^{1/2}-q^{-1/2})\omega_0^{(N)}}{k_+k_-\displaystyle\prod_{k=1}^{N}(-w_{0_1}^{(k)})(-w_{0_2}^{(k)})}W_{l+1}^{(N)}+\sum_{k=1}^{N}C_{-k+1}^{(N)}W_{k+l+1}^{(N)}+\epsilon_{(-)^{l+1}}^{(N)}=0\non\\
&&-\frac{(q^{1/2}-q^{-1/2})\omega_0^{(N)}}{k_+k_-\displaystyle\prod_{k=1}^{N}(-w_{0_1}^{(k)})(-w_{0_2}^{(k)})}G_{l+1}^{(N)}+\sum_{k=1}^{N}C_{-k+1}^{(N)}G_{k+l+1}^{(N)}=0\non\\
&&-\frac{(q^{1/2}-q^{-1/2})\omega_0^{(N)}}{k_+k_-\displaystyle\prod_{k=1}^{N}(-w_{0_1}^{(k)})(-w_{0_2}^{(k)})}\t G_{l+1}^{(N)}+\sum_{k=1}^{N}C_{-k+1}^{(N)}\t G_{k+l+1}^{(N)}=0\non.
\eeqa 
The above relations for the operators are a natural consequency in order to form closed algebra in finite dimensional case. 
The transfer matrix is drived as the following form:  
\beqa
  \label{traconserI}
     t^{(N)}(u)&=&Tr_0(K_+^{c}(u)K_-^{(N)}(u))\non\\
      &=&{\cal F}^{(N)}(u)+(u^2-u^{-2})(qu^2-q^{-1}u^{-2})
        \sum_{k=0}^{N-1}P_{-k}^{(N)}{\cal I}_{2k+1}^{(N)},
\eeqa
where the function ${\cal F}^{(N)}(u)$ is
\beqa
    {\cal F}^{(N)}(u)&=&(q^{1/2}+q^{-1/2})(\kappa^\ast\epsilon_+^{(N)}+\kappa\epsilon_{-}^{(N)})
                +(q^{1/2}u^2+q^{-1/2}u^{-2})(\kappa\epsilon_+^{(N)}+\kappa^\ast\epsilon_{-}^{(N)})\non\\
               &&\hspace{-1cm}+(q^{1/2}+q^{-1/2})(u^2-u^{-2})(qu^2-q^{-1}u^{-2})
                  \Big({\kappa_+}/{\Big(k_+\displaystyle\prod_{l=1}^{N}(-w^{(l)}_{0_1})\Big)}
                               +{\kappa_-}/{\Big(k_-\displaystyle\prod_{l=1}^{N}(-w^{(l)}_{0_2})\Big)}\Big) {\cal J}^{(N)}(u).
\eeqa
and the algebraic parts ${\cal I}_{2k+1}^{(N)}$ are derived as
\beq
      {\cal I}_{2k+1}^{(N)}=\left(\kappa W_{-k}^{(N)}+\kappa^\ast W_{k+1}^{(N)}
                             -\frac{\kappa_+}{k_+w^{(1)}_{0_1}} \t G_{k+1}^{(N)}
                               -\frac{\kappa_-}{k_-w^{(1)}_{0_2}} G_{k+1}^{(N)}  \right).
\eeq
By the construction for the transfer matrix,  ${\cal I}_{2k+1}^{(N)}$ should commute with each other.
It turns out that the commutativity of ${\cal I}_{2k+1}^{(N)}$ is ensured by the following algebraic relations
\beqa
  \label{commww}
    &&[W_{-k}^{(N)},W_{-l}^{(N)}]=0,\quad [W_{k+1}^{(N)},W_{l+1}^{(N)}]=0,\quad
      [G_{k+1}^{(N)},G_{l+1}^{(N)}]=0,\quad[\t G_{k+1}^{(N)},\t G_{l+1}^{(N)}]=0\\
  \noalign{\vskip 2mm}
    &&[W_{-k}^{(N)},W_{l+1}^{(N)}]=[W_{-l}^{(N)},W_{k+1}^{(N)}],\quad
      [W_{-k}^{(N)},G_{l+1}^{(N)}]=[W_{-l}^{N},G_{k+1}^{(N)}],\quad
      [W_{-k}^{(N)},\t G_{l+1}^{(N)}]=[W_{-l}^{N},\t G_{k+1}^{(N)}],\quad\\
  \noalign{\vskip 2mm}
    &&[W_{k+1}^{(N)},G_{l+1}^{(N)}]=[W_{l+1}^{N},G_{k+1}^{(N)}],\quad
      [W_{k+1}^{(N)},\t G_{l+1}^{(N)}]=[W_{l+1}^{N},\t G_{k+1}^{(N)}],\quad
      [G_{k+1}^{(N)},\t G_{l+1}^{(N)}]=[G_{l+1}^{N},\t G_{k+1}^{(N)}],\quad
\eeqa
with 
\beqa
    &&\hspace*{-10mm}[W_{k}^{(N)}-W_{-k}^{(N)},G_{l}^{(N)}]_q=[W_{l}^{(N)}-W_{-l}^{(N)},G_{k}^{(N)}]_q,\quad
      [W_{k}^{(N)}-W_{-k}^{(N)},\t G_{l}^{(N)}]_{q^{-1}}[W_{l}^{(N)}-W_{-l}^{(N)},\t G_{k}^{(N)}]_{q^{-1}},\\
  \noalign{\vskip 2mm}
    &&\hspace*{-10mm}[W_{-k+1}^{(N)}-W_{k+1}^{(N)},G_{l}^{(N)}]_{q^{-1}}=[W_{-l+1}^{(N)}-W_{l+1}^{(N)},G_{k}^{(N)}]_{q^{-1}},\ \ 
      [W_{-k+1}^{(N)}-W_{k+1}^{(N)},\t G_{l}^{(N)}]_{q}=[W_{-l+1}^{(N)}-W_{l+1}^{(N)},\t G_{k}^{(N)}]_{q},\\
  \noalign{\vskip 2mm}
    &&\hspace*{-10mm}\t G_{l}^{(N)}G_{k}^{(N)}-\t G_{k}^{(N)} G_{l}^{(N)}=\frac{(q^{1/2}+q^{-1/2})^3}{\t c}
      \left(\prod_{m=1}^{N}w_{0_1}^{(m)}w_{0_2}^{(m)}\right)\left([W_k^{(N)},W_{-l}^{(N)}]+[W_{-k}^{(N)},W_{l}^{(N)}]\right).
\eeqa
In fact these algebraic relations are verified by the mathematical induction with the reccursive expressions 
given in the appendix A. 

One can observe that the above algebraic equations possess the $q$-Doaln Grady relations even in the general $N$ case.
Some straightfoward calucurations show that $G_1^{(N)}=[W_1,W_0]_q$ and $\t G_1^{(N)}=[W_0,W_1]_q$.
From the initial conditions for the operator in the appedix A, we find the relations in the lowest order operators
\beqa
    &&W_2^{(N)}=W_0^{(N)}-\frac{1}{\rho_0^{(N)}}[W_1^{(N)},G_{1}^{(N)}]_{q^{-1}}\non\\
    &&W_{-1}^{(N)}=W_1^{(N)}+\frac{1}{\rho_1^{(N)}}[W_0^{(N)},G_{1}^{(N)}]_{q},
\eeqa
where $\rho^{(N)}_0=\rho^{(N)}_1=(q^{1/2}+q^{-1/2})^2k_-k_+\prod_{k=1}^{N}w_{-}^{(k)}w_{+}^{(k)}$.
Therefore one can easily obtain the $q$-Dolan Grady relations
\beqa
    &&[W_1^{(N)},[W_1^{(N)},[W_1^{(N)},W_0^{(N)}]_q]_{q^{-1}}]=\rho^{(N)}_0[W_1^{(N)},W_{0}^{(N)}],\quad\non\\
    &&[W_0^{(N)},[W_0^{(N)},[W_0^{(N)},W_1^{(N)}]_q]_{q^{-1}}]=\rho^{(N)}_1[W_0^{(N)},W_{1}^{(N)}],
\eeqa
from the algebraic relations in Eq. (\ref{commww}).

\section{Generalized McCoy-Wu model with open boundary conditions}
Here the generalized McCoy-Wu model with general open boundary conditions is considered as the most simplest model of our construction.
The Lax operators $L(u)$ and $\t L(u)$ are expressed by
\beq
    L^{(MW)}(u)=\left(\begin{array}{cc}
                q^{1/4}q^{\sigma_3/4}u-q^{-1/4}q^{-\sigma_3/4}u^{-1}&t^{-1/2}\t c\sigma_-\\
                                   t^{1/2}\t c\sigma_+    &q^{1/4}q^{-\sigma_3/4}u-q^{-1/4}q^{\sigma_3/4}u^{-1}\\ 
               \end{array}\right)\tau_g,
\eeq
\beq
    \t L^{(MW)}(u)=\tau_g^{-1}\left(\begin{array}{cc}
                q^{1/4}q^{\sigma_3/4}u^{-1}-q^{-1/4}q^{-\sigma_3/4}u&t^{-1/2}\t c\sigma_-\\
                                   t^{1/2}\t c\sigma_+    &q^{1/4}q^{-\sigma_3/4}u^{-1}-q^{-1/4}q^{\sigma_3/4}u\\ 
               \end{array}\right),
\eeq
through the fundamental spin-$\frac{1}{2}$ representation on a symmetric relization
\beq
\label{spinhalfrep}
    \t\tau_{1}^{\pm}=\mp q^{\mp1/4}q^{\mp\sigma_3/4}, \quad
    \t\tau_{2}^{\pm}=\mp q^{\mp 1/4}q^{\pm\sigma_3/4},\quad
    \t\tau_{12}=\t c\sigma_-,\quad 
    \t \tau_{21}=\t c\sigma_+,
\eeq
where $\sigma_{\pm}=(\sigma_1\pm i\sigma_2)/2$, and $\sigma_i,\ (i=1,2,3)$ are the Pauli matrices. 
Then the Casimir operators (\ref{Casimir}) is determined as
\beq
\label{spinhalfcas}
    w_\pm=q^{\mp 1/2},\quad w_{0_1}=w_{0_2}=-1,\quad w=(q+q^{-1}).
\eeq
In addition, we set $v_i=1 (i=1,\cdots N)$ to consider homogeneous spin chain model.
The Hamiltonian $H_{MW}$ of the system is obtained from  $\frac{d\ln t^{(MW)}(u)}{du}{\Big |}_{u\to1}$, where the transfer matrix is
\beq
    t^{(MW)}(u)={\rm Tr}_0\left(K_+^{c}(u)L^{(MW)}_N(u)\cdots L^{(MW)}_1(u)K_-^{c}(u)\t L^{(MW)}_1(u)\cdots
              \t L^{(MW)}_{N}(u)  \right).
\eeq
Differentiating $\ln t(u)$ with respect to the spectral parameter at $u=1$
\beq
    \frac{d\ln t^{(MW)}(u)}{du}{\Big |}_{u\to1}=
    \left(\frac{(q^{1/2}-q^{-1/2})}{(q^{1/2}+q^{-1/2})}
              +\frac{2N}{(q^{1/2}-q^{-1/2})}\Delta\right)I\!\!I^{(N)}
              +\frac{2}{(q^{1/2}-q^{-1/2})}H_{MW},
\eeq
we can obtain Hamiltonian for the generalized McCoy-Wu model with general open boundaies as follows:
\beqa
    \label{MWhami}
    H_{MW}&=&\sum_{k=1}^{N-1}\left(2t_{k+1}^{1/2}t_k^{-1/2}\sigma^{(k+1)}_{+}\sigma^{(k)}_{-}
                                 +2t_{k+1}^{-1/2}t_k^{1/2}\sigma^{(k+1)}_{-}\sigma^{(k)}_{+}
                                 +\Delta\sigma^{(k+1)}_3\sigma^{(k)}_3
                                 \right)\non\\
            &&+\frac{(q^{1/2}-q^{-1/2})}{(\kappa+\kappa^\ast)}
              \left(\frac{\kappa-\kappa^\ast}{2}\sigma^{(N)}_3
                         +2(q^{1/2}+q^{-1/2})(t^{1/2}_N\kappa_+\sigma_+^{(N)}+
                                               t^{-1/2}_N\kappa_-\sigma_-^{(N)}  )\right)\non\\
            &&+\frac{(q^{1/2}-q^{-1/2})}{(\epsilon_++\epsilon_-)}
              \left(\frac{\epsilon_+-\epsilon_-}{2}\sigma^{(1)}_3
                         +\frac{2}{(q^{1/2}-q^{-1/2})}(t^{1/2}_1 k_+\sigma_+^{(1)}+
                                               t^{-1/2}_1 k_-\sigma_-^{(1)}  )\right),           
\eeqa
where $\Delta=(q^{1/2}+q^{-1/2})/2$. The detailed analysis of the above derivation is given in the appendix B. 
It is clear that the Hamiltonian (\ref{MWhami}) for the untwisted limit, {\it i.e.} $t_i=1\ (i=1\cdots N)$, 
reduces to the XXZ open spin chain with general boundary conditions. 
By differentiating the transfer matrix $n$-times, mutually commuting quantities, 
which commute with the Hamiltonian, can be identified, although it seems that the method is much complicated. 

Alternatively, the transfer matrix by applying the Sklyanin's dressing method is described 
in terms of generators of the $q$-Onsager symmetry by taking the spin 1/2 representation (\ref{spinhalfrep}) and 
its related Casimir operators (\ref{spinhalfcas}).  
We denote the conserved quantities ${\cal I}_{2k+1}^{(N)}$ in the spin 1/2 representation as $I_{2k+1}^{(N)}$.
The Hamiltonian (\ref{MWhami}) is expressed by a linear combination of conserved quantities $I_{2k+1}^{(N)}$ from our construction for the transfer matrix. Thus all integrals of motion in this model are explicitly determined as $I_{2k+1}$, {\it i.e.} one can show that 
\beq
  [H_{MW},I_{2k+1}^{(N)}]=0, \quad(k=0,1,\cdots N-1).
\eeq 
In particular, the conserved quantity ${I}_1^{(N)}$ is given by
\beq
    I_{1}^{(N)}=\left(\kappa W_{0}^{(N)}+\kappa^\ast W_{1}^{(N)}
                             +\frac{\kappa_+}{k_+} \t G_{1}^{(N)}
                               +\frac{\kappa_-}{k_-} G_{1}^{(N)}  \right),
\eeq
where
\beqa
    &&W_0^{(N)}=(k_+t^{1/2}_N\sigma_+ +k_-t^{-1/2}_N\sigma_-)\otimes I\!\!I^{(N-1)}+q^{\sigma_3/2}\otimes W_0^{(N-1)}\non\\
    &&W_1^{(N)}=(k_+t^{1/2}_N\sigma_+ +k_-t^{-1/2}_N\sigma_-)\otimes I\!\!I^{(N-1)}+q^{-\sigma_3/2}\otimes W_1^{(N-1)}.
\eeqa
with the relations for $G_1^{(N)}=[W_1,W_0]_q$ and $\t G_1^{(N)}=[W_0,W_1]_q$. Instead of cosidering the eigenvalue problem $ H_{MW}$,
it may be read as the eigenvalue problem of the conserved quantity ${\cal I}_{1}^{(N)}$.

\section{Discussion}

We consider the ancestor model related to the twisted U$_q(sl_2)$ integrable models with integrable boundary conditions, which symmetric and nonsymmetric realizations of the algebra can generate several descendant lattice models without limiting procedures. The reflection equation and dual reflection equation for $K_\pm(u)$ does not depend on the twisted parameter $t$(or $\theta$). As expected that the same current algebra analyzed by Baseilhac and Shigechi is derived,
we observed that the transfer matrix of the model is generated by the $q$-Onsager symmetry, and 
we identified all of fundamental generators of the $q$-Onsager algebra in terms of the extended trigonometric Sklyanin algebra. Although there is no difference between the untwisted model and the twisted model in the reflection and dual reflection equations, their generators have explicit dependence on the twisted parameter. The twisted parameter dependence for the generatros shows that the new quantum integrable model with integrable boundary conditions were explicitly constructed.

We introduced the generalized McCoy-Wu model with general boundary condtions.
Mutually comutting quantities of the genelized McCoy-Wu spin chain model with general open boundaries 
were explicitly expressed in terms of abelian subalgebra of the $q$-Onsager algebra.
Therefore we confirmed that the model enjoys the $q$-Onsager symmetry.
To solve the spectrum problem is generally difficult in order that algebraic Bethe ansatz techniques have failed, because there is no reference state in the case of the general boundary conditions.  
Nice structure of the $q$-Onsager algebra may give useful information to solve this problem in similar to XXZ spin chain \cite{BK07}. However this analysis goes beyond the scope of this article. We plan to do this problem elsewhere.

\subsection*{Acknowledgements}
The author would like to thank J.H.H. Perk for communications and valuable comments.
The author would like to thank Maskawa Institute for Science and Culture at Kyoto Snagyo University 
that made it possible to complete this study.

\section{Appendix A} 
We give the recursive representation for the operators $ W_{-k}^{(N)},  W_{k+1}^{(N)},  G_{k+1}^{(N)}$ and $\t G_{k+1}^{(N)}$ in terms of the untiwsted extended Sklyanin algebra by dressing the $N-1$ dressed $K$ matrix again.
The direct method to find out $N$-dressed solution for boundary K-matrix finds that 
\beqa
    W_{-k}^{(N)}&=&\t\tau_1^{-}\t\tau_2^{+}\otimes(W^{(N-1)}_{k}-W^{(N-1)}_{-k})+\frac{w_0^{(N)}}{q^{1/2}+q^{-1/2}}I\!\!I\otimes W_k^{(N-1)}
    -\frac{(q^{-1/2}w_-^{(N)}v_N^2+q^{1/2}w_+^{(N)}v_N^{-2})}{(q^{1/2}+q^{-1/2})}I\!\!I\otimes W_{-k+1}^{(N-1)}\non\\
    &&+\frac{1}{k_+k_-(q^{1/2}+q^{-1/2})^2\displaystyle\prod_{k=1}^{N-1}(-w_{0_1}^{(k)})(-w_{0_2}^{(k)})}
    \left(k_+t_N^{1/2}v_N\t\tau_{21}\t\tau_1^{-}\otimes(\prod_{k=1}^{N-1}(-w_{0_1}^{(k)}))G_{k}^{(N-1)} \right.\non\\
    \noalign{\vskip -5mm}
    &&\hspace{9cm}\left.-k_-t_N^{-1/2}v_N^{-1}\t\tau_{12}\t\tau_2^{+}\otimes(\prod_{k=1}^{N-1}(-w_{0_2}^{(k)}))\t G_{k}^{(N-1)}   \right)\non\\
    &&+\frac{(q^{-1/2}w_-^{(N)}v_N^2+q^{1/2}w_+^{(N)}v_N^{-2})w_0^{(N)}}{w_{0_1}^{(N)}w_{0_2}^{(N)}(q^{1/2}+q^{-1/2})^2}W_{-k+1}^{(N)},\non
\eeqa
\beqa
    W_{k+1}^{(N)}&=&\t\tau_1^{+}\t\tau_2^{-}\otimes(W^{(N-1)}_{-k+1}-W^{(N-1)}_{k+1})+\frac{w_0^{(N)}}{q^{1/2}+q^{-1/2}}I\!\!I\otimes W_{-k+1}^{(N-1)}
    -\frac{(q^{-1/2}w_-^{(N)}v_N^2+q^{1/2}w_+^{(N)}v_N^{-2})}{(q^{1/2}+q^{-1/2})}I\!\!I\otimes W_{k}^{(N-1)}\non\\
    &&+\frac{1}{k_+k_-(q^{1/2}+q^{-1/2})^2\displaystyle\prod_{k=1}^{N-1}(-w_{0_1}^{(k)})(-w_{0_2}^{(k)})}
    \left(k_-t_N^{-1/2}v_N\t\tau_{12}\t\tau_2^{-}\otimes(\prod_{k=1}^{N-1}(-w_{0_2}^{(k)}))\t G_{k}^{(N-1)} \right.\non\\
    \noalign{\vskip -5mm}
    &&\hspace{9cm}\left.-k_+t_N^{1/2}v_N^{-1}\t\tau_{21}\t\tau_1^{+}\otimes(\prod_{k=1}^{N-1}(-w_{0_1}^{(k)}))G_{k}^{(N-1)}   \right)\non\\
    &&+\frac{(q^{-1/2}w_-^{(N)}v_N^2+q^{1/2}w_+^{(N)}v_N^{-2})w_0^{(N)}}{w_{0_1}^{(N)}w_{0_2}^{(N)}(q^{1/2}+q^{-1/2})^2}W_{k}^{(N)},\non
\eeqa
\beqa
   \hspace{-1cm} G_{k+1}^{(N)}&=&\frac{k_-t_{N}^{-1}\displaystyle\prod_{k=1}^{N}(-w_{0_2}^{(k)})}{k_+(q^{1/2}+q^{-1/2})\displaystyle\prod_{k=1}^{N-1}(-w_{0_1}^{(k)})}\t\tau_{12}^2\otimes \t G_{k}^{(N-1)}
       +\frac{w_{0_2}^{(N)}(q^{-1/2}v_N^2(\t\tau_1^-)^2+q^{1/2}v_N^{-2}(\t\tau_1^+)^2)}{(q^{1/2}+q^{-1/2})}\otimes G_{k}^{(N-1)}+w_{0_1}^{(N)}w_{0_2}^{(N)}I\!\!I\otimes G_{k+1}^{(N-1)}\non\\
    &&+k_-t_{N}^{-1/2}(q^{1/2}+q^{-1/2})\prod_{k=1}^{N}(-w_{0_2}^{(k)})
    \left(q^{1/2}v_N^{-1}\t\tau_{12}\t\tau_1^+\otimes(W_{-k+1}^{(N-1)}-W_{k+1}^{(N-1)})-q^{-1/2}v_N\t\tau_{12}\t\tau_1^-\otimes(W_{k}^{(N-1)}-W_{-k}^{(N-1)})\right)\non\\    &&+\frac{(q^{-1/2}w_-^{(N)}v_N^2+q^{1/2}w_+^{(N)}v_N^{-2})w_0^{(N)}}{w_{0_1}^{(N)}w_{0_2}^{(N)}(q^{1/2}+q^{-1/2})^2}G_{k}^{(N)}\non
\eeqa
and
\beqa
   \hspace{-1cm} \t G_{k+1}^{(N)}&=&\frac{k_+t_{N}\displaystyle\prod_{k=1}^{N}(-w_{0_1}^{(k)})}{k_+(q^{1/2}+q^{-1/2})\displaystyle\prod_{k=1}^{N-1}(-w_{0_2}^{(k)})}\t\tau_{21}^2\otimes G_{k}^{(N-1)}
       +\frac{w_{0_1}^{(N)}(q^{-1/2}v_N^2(\t\tau_2^-)^2+q^{1/2}v_N^{-2}(\t\tau_2^+)^2)}{(q^{1/2}+q^{-1/2})}\otimes \t G_{k}^{(N-1)}+w_{0_1}^{(N)}w_{0_2}^{(N)}I\!\!I\otimes \t G_{k+1}^{(N-1)}\non\\
    &&+k_+t_{N}^{1/2}(q^{1/2}+q^{-1/2})\prod_{k=1}^{N}(-w_{0_1}^{(k)})
    \left(q^{1/2}v_N^{-1}\t\tau_{21}\t\tau_2^+\otimes(W_{k}^{(N-1)}-W_{-k}^{(N-1)})-q^{-1/2}v_N\t\tau_{21}\t\tau_2^-\otimes(W_{-k+1}^{(N-1)}-W_{k+1}^{(N-1)})\right)\non\\    &&+\frac{(q^{-1/2}w_-^{(N)}v_N^2+q^{1/2}w_+^{(N)}v_N^{-2})w_0^{(N)}}{w_{0_1}^{(N)}w_{0_2}^{(N)}(q^{1/2}+q^{-1/2})^2}\t G_{k}^{(N)},\non
\eeqa
with the initial conditions
\beqa
    W_k^{(N)}\Big |_{k=0}\equiv 0,\quad W_{-k+1}^{(N)}\Big |_{k=0}\equiv 0
    ,\quad G_k^{(N)}\Big |_{k=0}=\t G_k^{(N)}\Big |_{k=0}\equiv\frac{k_+k_-(q^{1/2}+q^{-1/2})^2\displaystyle\prod_{k=1}^{N}(-w_{0_1}^{(k)})(-w_{0_2}^{(k)})}{(q^{1/2}-q^{-1/2})}I\!\!I^{(N)}
\eeqa
and
\beqa
    W_{-l}^{(0)}=\epsilon_{(-)^l}^{(0)},\quad
     W_{l+1}^{(0)}=\epsilon_{(-)^{l+1}}^{(0)},\quad
     G_{l+1}^{(0)}=\t G_{l+1}^{(0)}\equiv\epsilon_+^{(0)}\epsilon_-^{(0)}(q^{1/2}-q^{-1/2}),\quad(l=0,1,2,\cdots).
\eeqa

\section{Appendix B}
In the Appendix B, the Hamiltonian $H_{MW}$ of the generalized McCoy-Wu model \cite{MW,Kun} with general open boundaries is explicitly deriven, because it is some complications compared with one of the XXZ opens spin chain with general boundaries.
We can perform to differentiate $\ln t(u)$ with respect to the spectral parameter $u$ at $u=1$ by using the Libnitz rule:
\beqa
    &&\hspace{-1cm}\frac{d\ln t^{(MW)}(u)}{du}{\Big |}_{u\to1}\non\\
    &=&\frac{1}{t^{(MW)}(1)}\left[{\rm Tr}_0\left(\dot{K}_+^{c}(1)L^{(MW)}_N(1)\cdots L^{(MW)}_1(1)K_-^{c}(1)\t L^{(MW)}_1(1)\cdots
              \t L^{(MW)}_{N}(1)  \right)\right.\non\\
              &&\hspace{1cm}+{\rm Tr}_0\left({K}_+^{c}(1){\dot L}^{(MW)}_N(1)\cdots L^{(MW)}_1(1)K_-^{c}(1)\t L^{(MW)}_1(1)\cdots
              \t L^{(MW)}_{N}(1)\right)\non\\
              &&\hspace{1cm}+\sum_{i=1}^{N-1}{\rm Tr}_0\left({K}_+^{c}(1)L^{(MW)}_N(1)\cdots {\dot L}^{(MW)}_i(1)\cdots L^{(MW)}_1(1)K_-^{c}(1)\t L^{(MW)}_1(1)\cdots
              \t L^{(MW)}_{N}(1)\right)\non\\
              &&\hspace{1cm}+{\rm Tr}_0\left({K}_+^{c}(1){L}^{(MW)}_N(1)\cdots L^{(MW)}_1(1){\dot K}_-^{c}(1)\t L^{(MW)}_1(1)\cdots
              \t L^{(MW)}_{N}(1)\right)\non\\
              &&\hspace{1cm}+\sum_{i=1}^{N-1}{\rm Tr}_0\left({K}_+^{c}(1){L}^{(MW)}_N(1)\cdots L^{(MW)}_1(1){K}_-^{c}(1)\t L^{(MW)}_1(1)\cdots{\dot{\t L}}^{(MW)}_i(1)\cdots
              \t L^{(MW)}_{N}(1)\right)\non\\
             &&\left.\hspace{1cm}+{\rm Tr}_0\left({K}_+^{c}(1){L}^{(MW)}_N(1)\cdots L^{(MW)}_1(1){K}_-^{c}(1)\t L^{(MW)}_1(1)\cdots
              {\dot{\t L}}^{(MW)}_{N}(1)\right) \right]\non.
\eeqa
It is convinient for the derivetion that $L^{(MW)}(1)$ and $\t L^{(MW)}(1)$ are expressed as
\beqa
    &&L^{(MW)}_i(1)=\t c\left(\begin{array}{cc}
                            \displaystyle\frac{1+\sigma^{(i)}_3}{2}+t_i^{-1/2}\frac{1-\sigma^{(i)}_3}{2}&0\\
                            0&\displaystyle t_i^{1/2}\frac{1+\sigma^{(i)}_3}{2}+\frac{1-\sigma^{(i)}_3}{2}
                            \end{array}\right){\cal P}_{0,i}\tau_g^{(i)},\non\\
  \noalign{\vskip 2mm}
    &&\t L^{(MW)}_i(1)=\t c(\tau_g^{(i)})^{-1}{\cal P}_{0,i}\left(\begin{array}{cc}
                            \displaystyle \frac{1+\sigma^{(i)}_3}{2}+t_i^{1/2}\frac{1-\sigma^{(i)}_3}{2}&0\\
                            0&\displaystyle t_i^{-1/2}\frac{1+\sigma^{(i)}_3}{2}+\frac{1-\sigma^{(i)}_3}{2}
                            \end{array}\right),\non
\eeqa
where ${\cal P}_{i,j}$ denotes the permutation operator with properties
\beq
    {\cal P}_{i,j}={\cal P}_{j,i},\quad {\cal P}_{i,j}^2=1,\quad {\cal P}_{i,j}F_{j,k}=F_{i,k}{\cal P}_{i,j}\quad(k\ne i,j) \non
\eeq 
for any operator $F_{j,k}$ acting on lattice sites $j$ and $k$. The first term is 
\beqa
    &&\frac{1}{t^{(MW)}(1)}{\rm Tr}_0\left(\dot{K}_+^{c}(1)L^{(MW)}_N(1)\cdots L^{(MW)}_1(1)K_-^{c}(1)\t L^{(MW)}_1(1)\cdots
              \t L^{(MW)}_{N}(1)  \right)\non\\
    &&\hspace{-5mm}=\frac{1}{(q^{1/2}+q^{-1/2})(\kappa+\kappa^\ast)} {\rm Tr}_0\left(\dot{K}_+^{c}(1)\right)
                   =\frac{q^{1/2}-q^{-1/2}}{q^{1/2}+q^{-1/2}},\non
\eeqa
with the relation
\beq
    t^{(MW)}(1)=\t c^{2N}(q^{1/2}+q^{-1/2})(\epsilon_++\epsilon_-)(\kappa+\kappa^\ast)I\!\!I. \non
\eeq
The second term contributes the left boundary term of $H_{MW}$:
\beqa
    &&\frac{1}{ t^{(MW)}(1)}{\rm Tr}_0\left({K}_+^{c}(1){\dot L}^{(MW)}_N(1)\cdots L^{(MW)}_1(1)K_-^{c}(1)\t L^{(MW)}_1(1)\cdots
              \t L^{(MW)}_{N}(1)\right)\non\\
    &&\hspace{-8mm}=\frac{1}{2(q-q^{-1})(\kappa+\kappa^\ast)}
    {\rm Tr}_0\left({K}_+^{c}(1)\left(\begin{array}{cc}
                                 \xi^2+\eta^2\sigma_3^{(N)}&0\\
                                  0&\xi^2-\eta^2\sigma_3^{(N)}
                                \end{array}\right)
                                \left(\begin{array}{cc}
                                 \frac{1+\sigma_3^{(N)}}{2}&t^{-1/2}_N\sigma_-^{(N)}\\
                                  t^{1/2}\sigma_{+}^{(N)}&\frac{1-\sigma_3^{(N)}}{2}     
                                \end{array}\right)\right)\non\\
    &&\hspace{-8mm}=\frac{1}{2\t c(\kappa+\kappa^\ast)}
    {\rm Tr}_0\left(\left(\begin{array}{cc}
                                 q^{1/2}\kappa+q^{-1/2}\kappa^\ast&\kappa_+(q^{1/2}+q^{-1/2})^2\t c\\
                                  \kappa_-(q^{1/2}+q^{-1/2})^2\t c&q^{1/2}\kappa^\ast+q^{-1/2}\kappa
                                \end{array}\right)
                                \left(\begin{array}{cc}
                                  1+\sigma_3^{(N)}&\frac{4t^{-1/2}_N}{q^{1/2}+q^{-1/2}}\sigma_-^{(N)}\\
                                  \frac{4t^{1/2}_N}{q^{1/2}+q^{-1/2}}\sigma_{+}^{(N)}&1-\sigma_3^{(N)}     
                                \end{array}\right)\right)\non\\
    &&\hspace{-8mm}=\frac{\Delta}{(q^{1/2}-q^{-1/2})}
        +\frac{1}{(\kappa+\kappa^\ast)}
        \left(\frac{\kappa-\kappa^\ast}{2}\sigma_3^{(N)}
                                       +2(q^{1/2}+q^{-1/2})(t_N^{1/2}\kappa_+\sigma_+^{(N)}
                                                             +t_N^{-1/2}\kappa_-\sigma_-^{(N)}
)\right),\non                        
\eeqa
where we used notations $\xi=(q^{1/4}+q^{-1/4}),\ \eta=(q^{1/4}-q^{-1/4})$. Similarily, we find that the last term is equal to the second one. To obtain the third term, a key point relation is the followings:
\beqa
   &&\hspace{-1cm}L_{i+1}^{(MW)}(1){\dot L}_{i}^{(MW)}(1){\t L}_{i}^{(MW)}(1){\t L}_{i+1}^{(MW)}(1)\non\\
    &=&\frac{\t c}{2}L_{i+1}^{(MW)}(1)
            \left(\begin{array}{cc}
                    \xi^2+\eta^2\sigma^{(i)}_3&0\\
                    0&\xi^2-\eta^2\sigma^{(i)}_3\\
                  \end{array}\right)
           \left(\begin{array}{cc}
                    \frac{1+\sigma^{(i)}_3}{2}&t^{-1/2}_i\sigma_-^{(i)}\\
                    t_i^{1/2}\sigma_+^{(i)}&\frac{1-\sigma^{(i)}_3}{2}\\
                  \end{array}\right){\t L}_{i+1}^{(MW)}(1)\non\\
    &=&\frac{\t c^3(q^{1/2}+q^{-1/2})}{2} \left(\begin{array}{cc}
                    \frac{1+\sigma^{(i+1)}_3}{2}+\frac{t_{i+1}^{-1/2}(1-\sigma_3^{(i+1)})}{2}&0\\
                    0&\frac{t^{1/2}_{i+1}(1+\sigma^{(i)}_3)}{2}+\frac{1-\sigma^{(i)}_3}{2}\\
                  \end{array}\right){\cal P}_{0,{i+1}}\times\non\\
    &&\hspace{2cm}\times\left(\begin{array}{cc}
                  1+\sigma_3^{(i)}&\frac{4t_i^{-1/2}}{q^{1/2}+q^{-1/2}}\sigma_-^{(i)}\\
                  \frac{4t_i^{1/2}}{q^{1/2}+q^{-1/2}}\sigma_+^{(i)}&1-\sigma_3^{(i)}
                  \end{array}\right){\cal P}_{0,{i+1}}\left(\begin{array}{cc}
                    \frac{1+\sigma^{(i+1)}_3}{2}+\frac{t_{i+1}^{1/2}(1-\sigma_3^{(i+1)})}{2}&0\\
                    0&\frac{t^{-1/2}_{i+1}(1+\sigma^{(i)}_3)}{2}+\frac{1-\sigma^{(i)}_3}{2}\\
                  \end{array}\right)\non\\
   &=&\frac{\t c^3(q^{1/2}+q^{-1/2})}{2} \left(\begin{array}{cc}
                    \frac{1+\sigma^{(i+1)}_3}{2}+\frac{t_{i+1}^{-1/2}(1-\sigma_3^{(i+1)})}{2}&0\\
                    0&\frac{t^{1/2}_{i+1}(1+\sigma^{(i)}_3)}{2}+\frac{1-\sigma^{(i)}_3}{2}\\
                  \end{array}\right)\times\non\\
    &&\hspace{2cm}\times\left(1\otimes1+\sigma_3^{(i+1)}\otimes\sigma_{3}^{(i)}
                  +\frac{4t_i^{-1/2}}{q^{1/2}+q^{-1/2}}\sigma_+^{(i+1)}\otimes\sigma_{-}^{(i)}
                  +\frac{4t_i^{1/2}}{q^{1/2}+q^{-1/2}}\sigma_-^{(i+1)}\otimes\sigma_{+}^{(i)}\right)\times\non\\
    &&\hspace{3cm}\times\left(\begin{array}{cc}
                    \frac{1+\sigma^{(i+1)}_3}{2}+\frac{t_{i+1}^{1/2}(1-\sigma_3^{(i+1)})}{2}&0\\
                    0&\frac{t^{-1/2}_{i+1}(1+\sigma^{(i)}_3)}{2}+\frac{1-\sigma^{(i)}_3}{2}\\
                  \end{array}\right)\non\\
    &&=\frac{\t c^3(q^{1/2}+q^{-1/2})}{2}\left(1\otimes1+\sigma_3^{(i+1)}\otimes\sigma_{3}^{(i)}
                  +\frac{4t_{i+1}^{1/2}t_i^{-1/2}}{q^{1/2}+q^{-1/2}}\sigma_+^{(i+1)}\otimes\sigma_{-}^{(i)}
                  +\frac{4t_{i+1}^{-1/2}t_i^{1/2}}{q^{1/2}+q^{-1/2}}\sigma_-^{(i+1)}\otimes\sigma_{+}^{(i)}\right)
                  \left(\begin{array}{cc}1&0\\0&1\end{array}\right)\non
\eeqa
Therefore the third term leads to the bulk interaction term:
\beqa
    &&\frac{1}{t^{(MW)}(1)}\sum_{i=1}^{N-1}{\rm Tr}_0\left({K}_+^{c}(1)L^{(MW)}_N(1)\cdots {\dot L}^{(MW)}_i(1)\cdots L^{(MW)}_1(1)K_-^{c}(1)\t L^{(MW)}_1(1)\cdots
              \t L^{(MW)}_{N}(1)\right)\non\\
   &=&\frac{(N-1)\Delta}{(q^{1/2}-q^{-1/2})}
   +\frac{1}{q^{1/2}-q^{-1/2}}\sum_{i=1}^{N-1}
   \left(2t_{i+1}^{1/2}t_{i}^{-1/2}\sigma_+^{(i+1)}\sigma_-^{(i)}
          +2t_{i+1}^{-1/2}t_{i}^{1/2}\sigma_-^{(i+1)}\sigma_+^{(i)}+\Delta\sigma_3^{(i+1)}\sigma_3^{(i)}
   \right),\non
\eeqa
and the fifth term gives the same contribution as the third one. By using the relation
\beqa
    &&L_1^{(MW)}(1){\dot K}_-^{c}(1){\t L}_1^{(MW)}(1)\non\\
    &&=\t c^2\left(\begin{array}{cc}
                   \frac{1+\sigma^{(1)}_3}{2}+\frac{t^{-1/2}_1(1-\sigma_3^{(1)})}{2}&0\\
                   0& \frac{t_1^{1/2}(1+\sigma^{(1)}_3)}{2}+\frac{1-\sigma_3^{(1)}}{2}
                   \end{array}\right){\cal P}_{01}
             \left(\begin{array}{cc}
                   \epsilon_{+}-\epsilon_-&4k_+/\t c\\
                   4k_-/\t c&\epsilon_{-}-\epsilon_+      
             \end{array}\right)\times\non\\
     &&\hspace{3cm}\times{\cal P}_{01}
             \left(\begin{array}{cc}
                   \frac{1+\sigma^{(1)}_3}{2}+\frac{t^{1/2}_1(1-\sigma_3^{(1)})}{2}&0\\
                   0& \frac{t_1^{-1/2}(1+\sigma^{(1)}_3)}{2}+\frac{1-\sigma_3^{(1)}}{2}
                   \end{array}\right)\non\\
    &&=\t c^2\left(\begin{array}{cc}
                   \frac{1+\sigma^{(1)}_3}{2}+\frac{t^{-1/2}_1(1-\sigma_3^{(1)})}{2}&0\\
                   0& \frac{t_1^{1/2}(1+\sigma^{(1)}_3)}{2}+\frac{1-\sigma_3^{(1)}}{2}
                   \end{array}\right)
       \left((\epsilon_+-\epsilon_-)\sigma_3^{(1)}
                        +\frac{4}{\t c}(k_+\sigma_{+}^{(1)}+k_-\sigma_{-}^{(1)})\right)\otimes\non\\
    &&\hspace{3cm}\otimes\left(\begin{array}{cc}
                   \frac{1+\sigma^{(1)}_3}{2}+\frac{t^{1/2}_1(1-\sigma_3^{(1)})}{2}&0\\
                   0& \frac{t_1^{-1/2}(1+\sigma^{(1)}_3)}{2}+\frac{1-\sigma_3^{(1)}}{2}
                   \end{array}\right)\non\\
    &&=\t c^2\left((\epsilon_+-\epsilon_-)\sigma_3^{(1)}
                        +\frac{4}{\t c}(t_1^{1/2}k_+\sigma_{+}^{(1)}+t_1^{-1/2}k_-\sigma_{-}^{(1)})\right)
               \left(\begin{array}{cc}1&0\\
                                       0&1
                     \end{array}\right),\non
\eeqa
the middle term can be identified with the right boundary term:
\beqa
    &&\hspace{-1cm}\frac{1}{t^{(MW)}(1)}{\rm Tr}_0\left({K}_+^{c}(1){L}^{(MW)}_N(1)
                               \cdots L^{(MW)}_1(1){\dot K}_-^{c}(1)\t L^{(MW)}_1(1)\cdots
                                                               \t L^{(MW)}_{N}(1)\right)\non\\
    &=&\frac{2}{\epsilon_++\epsilon_-}
    \left(\frac{\epsilon_+-\epsilon_-}{2}\sigma_3^{(1)}
                  +\frac{2}{q^{1/2}-q^{-1/2}}(t_1^{1/2}k_+\sigma_{+}^{(1)}+t_1^{-1/2}k_-\sigma_{-}^{(1)})\right).\non       
\eeqa
Gathering all of these results permits us to derive the form of Eq.(\ref{MWhami}).

\end{document}